\documentstyle[fleqn]{article}
\def\BE{\begin{equation}}
\def\EE{\end{equation}}
\def\BEA{\begin{eqnarray}}
\def\EEA{\end{eqnarray}}
\def\BC{\begin{center}}
\def\EC{\end{center}}

\def\tr{{\rm tr}}
\def\re{{\rm Re}}

\def\half{\frac{1}{2}}

\begin{document}

\title{Variational Approach to Real-Time Evolution of
	Yang-Mills Gauge Fields on the Lattice}
\author{
	C. Gong and B. M\"uller\\
        Physics Department, P. O. Box 90305, Duke University\\
        Durham, NC 27708-0305, U.S.A.\\
	and\\
	T. S. Bir\'o\\
	Institut f\"ur Theoretische Physik\\
	Justus-Liebig-Universit\"at\\
	Heinrich-Buff-Ring 16, D-6300 Giessen, Germany\\
	}
\maketitle
\begin{abstract}
Applying a variational method to a Gaussian wave ansatz,
we have derived
a set of semi-classical evolution equations for SU(2) lattice gauge
fields, which
take the classical form in the limit of a
vanishing width of the Gaussian wave packet.
These equations are used to study the quantum effects on the classical
evolutions of the lattice gauge fields.
\end{abstract}
\newpage

\section{ Introduction}

Chaotic behavior in non-abelian gauge fields is first shown for
the spatially homogenous potential configurations \cite{matinyan} and
later also for the spherically sysmmetric fields \cite{matinyan2}.
Recently a real time evolution method is introduced to
study general non-abelian gauge fields on a 3 dimensional lattice,
where chaoticity is shown by obtaining a positive
maximal Lyapunov exponent
\cite{muller}.
The maximal Lyapunov exponent is
found to be related to the damping rate of gluons at rest, an infrared
property of the quantized system. It can be argued that in the high
temperature limit the properties associated with the
infrared degrees of freedom in the gauge system are
classical at the leading order. But it is still reasonable
to ask how reliable these classical results are and what
kind of effects the quantum
corrections might bring about.

A classical dynamical system can be studied in two ways. One is
by tracing single particle trajectories using Newton's
equation. The other is by studying the phase space density evolution
based on Liouville's equation.
The links between classical and quantum physics have been
studied in both formalisms for a long time.
One of these links is the Wigner function which
is the counterpart of the
phase space distribution in classical dynamics \cite{wigner}.
The driving force of its evolution can be
separated into classical and quantum parts \cite{remler}. Unfortunately
the variables in
the lattice gauge theory are defined on
a compact group manifold with a non-trivial geometry. The Wigner function
on a compact position space
is defined on a discrete momentum space \cite{hannay},
rendering the classical limit rather subtle.
Therefore
we want to explore a more direct analogy to
a classical trajectory in
quantum physics, namely the evolution of a narrow wave packet.
In order to resemble a classical trajectory as close
as possible, the width of the wave packet in
coordinate space and that in momentum space shall satisfy
the uncertainty relation most efficiently,
which suggests that we shall choose
a Gaussian.

In this paper, as our first attack on the problem of quantum effects on
the classical evolution of a gauge system, we apply the
variational method to Gaussian wave packets \cite{biro,kramer}.
The basic idea is to parametrize a wave packet,
usually a Gaussian one, by a small number of parameters.
Then the time dependent variational principle is used
to derive a set of evolution
equations for these parameters, which have a symplectic structure
like Hamiltonian equations. If the parameters
are chosen in such a way that some of them
resemble the variables in the corresponding classical system,
then the coupling between the semi-classical time evolution of these
parameters and that of all others can be interpreted as
quantum corrections to the classical evolution.
This method has been used to treat the scattering problem of
$\alpha$ particles \cite{saraceno},
and recently it also has been applied to
the many body problem of interacting fermion systems
\cite{feldmeier,coriano}.
In the following we will first introduce the method. Then it will
be applied to the pure SU(2) lattice gauge system. The semi-classical
Hamiltonian and the equations of motion will be derived and their
solutions will be studied numerically.

\section{ Method}

We begin by considering a quantum system described by a Hamiltonian
$H$. The dynamics of the
system is contained in the Schr\"odinger equation
\BE
i\hbar\partial_{t} |\Phi(t)\rangle = H |\Phi(t)\rangle.
\EE
We solve this equation by the time-dependent variational method \cite{kramer}
as follows.
A normalized trial wavefunction $|\Phi(q)\rangle$,
where $q$ denotes a set of variational parameters,
must satisfy the following variational principle
\BE
\delta \int_{t1}^{t2} dt\langle \Phi(q(t))|(i\hbar\partial_{t}-
                H)|\Phi(q(t))\rangle =0.
\EE
Among these parameters
some have definite meanings in the classical limit, such as the classical
momenta and positions.
Provided that enough parameters are chosen to span the whole Hilbert
space, (2) can solve (1) exactly.
Another case, when this is possible, is if we
choose the parameters so clever that the exact time
evolution of the initial state admixes only states
from that part of the Hilbert space which is spanned by these trial states.
But generally (2) is only an approximation to (1).
Performing the variation in the usual way we can derive
an effective Lagrangian
\BE
L = \dot{q}_i z_i(q) - {\cal H}(q)
\EE
with
\BE
z_i(q) = \frac{i}{2}
\left( < \Phi | \frac{\partial \Phi}{\partial q_i} >
 - < \frac{\partial \Phi}{\partial q_i} | \Phi > \right)
\EE
and
\BE
{\cal H}(q) = < \Phi | H | \Phi>.
\EE
In the equations of motion,
\BE
A_{ij}(q) \dot{q}_j = \frac{\partial {\cal H}}{\partial q_i},
\EE
the antisymmetric coefficient,
\BE
A_{ij}(q) =
i \left. \left( \frac{\partial}{\partial q_i'} \frac{\partial}{\partial q_j}
- \frac{\partial}{\partial q_i} \frac{\partial}{\partial q_j'} \right)
\ln \left< \Phi(q') | \Phi(q) \right> \right|_{q'=q},
\EE
occurs, which satisfies the Jacobi identity as well,
\BE
\frac{\partial A_{ij}}{\partial q_k} + {\rm cyclic \quad permutations} = 0.
\EE
Since $A_{ij}$ is antisymmetric, the expectation value of the energy,
${\cal H}$, is conserved,
\BE
\frac{d{\cal H}}{dt} = \dot{q}_i A_{ij} \dot{q}_j = 0.
\EE
Other conservation laws related to symmetries of the quantum system
can be derived accordingly.

\par
Suppose that $G$ is a generator of an infinitesimal symmetry transformation
related to a conserved charge, i.e. it commutes with the Hamilton operator
\BE
[H,G] = 0.
\EE
Denote the corresponding 'classical operator' by $G_0$, whose action
on the respective classical parameter $x_{0}$ is identical to that of the
'quantum operator' $G$ on the variable $x$
\BE
G: x \rightarrow f(x) \qquad \qquad G_0: x_{0} \rightarrow f(x_{0}).
\EE
We call our trial wave function covariant under the transformation $G$ if
it is invariant under the joint operation of $G$ and $G_0$,
\BE
(G+G_0) | \Phi > = 0.
\EE
Such wave functions lead to a conserved expectation value
\BE
\frac{d {\cal G}}{dt} = 0,
\EE
with
\BE
{\cal G} = < \Phi | G | \Phi>.
\EE

To prove this, let us consider
an infinitesimal symmetry transformation generated
by $G_0$. From the variational principle (2), we have
\BE
\int \! dt \left( < \delta \Phi | (i\hbar \partial_t - H) | \Phi > +
< \Phi | (i\hbar \partial_t - H) | \delta \Phi > \right) = 0,
\EE
with
\BEA
| \delta \Phi > = i \delta \alpha G_0 | \Phi > \nonumber \\
\nonumber \\
< \delta \Phi | = - i \delta \alpha < \Phi | G_0,
\EEA
where $\delta \alpha$ is a small {\em real} parameter.
\par
The above integrand can be casted into the form
\BE
- \delta \dot{\alpha} < \Phi | G_0 | \Phi >
- \delta \alpha < \Phi | \dot{G}_0 | \Phi >
- i \delta \alpha < \Phi | [G_0,H] | \Phi >.
\EE
Now using $G_0 | \Phi > = - G | \Phi >$ the commutator term vanishes,
because $G$ commutes with $H$. The rest can be splitted into a total
time derivative, which we ignore in the variational integral, and
into terms proportional with $\delta \alpha$. Finally we get
\BE
\int \! dt \delta \alpha
\left( < \dot{\Phi} | G_0 | \Phi > + < \Phi | G_0 | \dot{\Phi} > \right)
= 0.
\EE
Using again $G_0 | \Phi > = - G | \Phi >$ and the fact that
$\dot{G} = 0$,
we arrive at
\BE
- \int \! dt \delta \alpha
\left( < \dot{\Phi} | G | \Phi > + < \Phi | \dot{G} | \Phi > +
        < \Phi | G | \dot{\Phi} > \right) = 0,
\EE
which, for $\delta \alpha$ being arbitrary, is equivalent to
\BE
\frac{d}{dt} < \Phi | G | \Phi > = 0.
\EE
We intend to use this property of the time-dependent variational
principle in the following in order to ensure that Gauss's Law is
satisfied on the average during the time evolution of Yang-Mills
systems.

\section{ SU(2) Lattice Gauge Theory}

\subsection{ Variational Wavefunctions}

Now we apply this method to
lattice gauge theory.
Here we restrict ourselves to SU(2) theory for simplicity.
The lattice gauge Hamiltonian is \cite{chin}
\begin{equation}
H=\frac{g^{2}}{a}\left(\sum_{l}\frac{1}{2} E^{a}_{l} E^{a}_{l}+\frac{4}{g^{4}}
 \re \sum_{p}(1-\half \tr U_{p})\right),
\end{equation}
where the first term corresponds to the electric and the
second to the magnetic energy.
In the classical limit the pure gauge theory is scale invariant. The
coupling constant $g^{2}$ and the lattice spacing $a$ can be
scaled out entirely and we are left with no free parameter except the
total energy (or temperature). In the semiclassical approach we
want to consider the quantum corrections to classical trajectories
so Planck's constant $\hbar$ comes in naturally. After
making the following scaling transformation,
\BEA
H^{'}  = & ag^{2}H,  \nonumber \\
t^{'}  = & t/a, \nonumber \\
E^{'}  = & g^{2}E, \nonumber  \\
\hbar^{'}  = & g^{2}\hbar,
\EEA
the variables on the left become
dimensionless numbers. In the following all quantities
are understood to be the scaled ones unless otherwise
stated explicitly. The equations all stay the same except
that we shall put $a=1$ and $g=1$.
The scaled parameter $\hbar=4\pi\alpha_{s}$,
where $\alpha_{s}=g^{2}\hbar/4\pi$
is the dimensionless coupling constant in the gauge theory,
enters our formalism in two ways. First it appears in the variational
principle,
\BE
\delta \int_{t1}^{t2} \langle \Phi|(i\hbar\partial_{t}-
                H)|\Phi \rangle =0.
\EE
Secondly it also enters
the commutation relation,
\BE
[E^{a}_{l},A^{b}_{m}]=-i\hbar \delta^{ab}\delta_{lm}.
\EE
The explicit appearance of coupling constant in our calculation fixes
the scale of quantum effects relative to classical effects.

To proceed we need to choose a trial wavefunction. In the
strong coupling and zero temperature limit, the ground
state wave function is an expansion of
plaquette variables $tr U_{p}$ \cite{chin}.
It is separable for different plaquettes,
but this choices is
not very practical in our context. First we want the trial function
to have the feature that the averages can be performed in closed
analytical expressions.
A plaquette separable wavefunction
is more difficult in this aspect than
a link separable one (see below).
The second reason is that we are mainly interested in the high temperature
limit where the system can be treated classically. In this
limit the former is not a priori better
justified than the latter. Moreover, since in the classical
limit we deal with link variables rather than plaquette variables,
a link separable wavefunction
allows to make a more direct connection to classical dynamics.
We choose the following Gaussian ansatz,
\BE
\Phi[U_{l}]=\prod_{l}\phi_{l}(U_{l})
	=\prod_{l}
        \frac{1}{\sqrt{N_{l}}} \exp\left(\frac{b_{l}}{2} \tr (U_{l}U_{l0}^{-1})
                -\frac{1}{\hbar}\tr (E_{l0}U_{l}U_{l0}^{-1}) \right).
\EE
The $U_{l0}$ and $E_{l0}$ are
parameters of the wavefunction, which correspond to classical link variables
and left electric fields, respectively.
Physically they denote the center of the coordinates and momenta of the
Gaussian wave function.
The $U_{l}$'s and $U_{l0}$'s are group elements of SU(2),
whereas the $E_{l0}$ are elements
of the su(2) algebra. The complex parameter $b=v+iw$ controls the width,
or more explicitely, $\sqrt{v}$ is the inverse of the width.
$N_{l}$ is a normalization constant to be specified later.

The form (25) is chosen so that it is covariant under the
gauge transformations, i.e. invariant under the simultaneous gauge
transformation on both the quantum variables $U_{l}$ and the
classical parameters $U_{l0}$ and $E_{l0}$,
\BE
U_{l}' = g_{L} U_{l} g^{-1}_{R} \qquad U_{l0}' = g_{L} U_{l0} g^{-1}_{R} \qquad
{\rm and} \qquad E_{l0}' = g_{L} E_{l0} g^{-1}_{L},
\EE
where $g_{L(R)}$ is a gauge transformation at the site left(right)
to the link $l$.
This leads to the
desirable result that Gauss's law is conserved on average.

We note here that color confinement
requires that all physical states shall be color singlet,
while the trial wavefunction we use here is not gauge invariant.
Is this a contradiction? The answer is yes in the sense that our
choice does not correspond directly to a physical state.
But here we have two arguments to support our choice.
First we note that a covariant discription is
as close as we can get in the classical limit where we are always
using color octet variables $U_{l}$ and $E_{l}$ to define a gauge
configuration.
Had we started from a color singlet trial wavefunction and wanted
to derive classical equations of motion for averages of these variables,
we would always get vanishing expectation
values because of the Wigner-Eckart theorem.
The second argument is that
given a gauge covariant wavefunction we can easily project out its singlet
part. So the knowledge of
the former can help to understand the properties of a gauge invariant
state.

In the continuum limit, this wavefunction formally coincides with a
Gaussian in the vector potential $A(x)$,
\BEA
\Phi \left[(A(x)\right] &=&
\frac{1}{\sqrt{N}} \exp \left[ \int d^{3}x
	\sum_{a}\left(- \frac{b(x)}{8}(A^{a}(x)-A^{a}_{0}(x))^{2}+
                \frac{i}{\hbar}E_{0}(x)^{a}A^{a}(x) \right) \right]
		\nonumber \\.
		& &
\EEA
The wavefunction (25) is normalized on each link with
\BE
N(v)= 2\pi^{2} I_{1}(2 v)/v,
\EE
where $I_{n}$ denotes the modified  Bessel function.
Wavefunctions with different parameters are not orthogonal to each other.
They form a set of overcomplete basis states of the Hilbert space
for each link, with the completeness relation
\BE
\int dU_{0}d^{3}E_{0} \Phi(E_{0},U_{0},b)\Phi(E_{0},U_{0},b)=1,
\EE
for arbitrary $b$. This permits us to deal with the complex width
parameter $b$ in two ways.
We could define a measure $\mu(b)$ which satisfies
\BE
\int d\mu(b)=1,
\EE
so that we formally have complete closure relation over all parameters
\BE
\int dU_{0}d^{3}E_{0}d\mu(b) \Phi(E_{0},U_{0},b)\Phi(E_{0},U_{0},b)=1.
\EE
The exact form of $\mu(b)$ is not known, but could be
obtained with numerical techniques.
On the other hand we can treat $b$ as
an additional variational parameter. We will adopt this point of view here.

The relation between our classical parameters and the averaged quantities
are obtained as
\BEA
\langle E^{a}\rangle =f(v)E^{a}_{0}, \nonumber \\
\langle E^{a}_{R} \rangle =f(v)E^{a}_{R0}, \nonumber \\
\langle U \rangle = f(v)U_{0},
\EEA
where $E_{R}^{a}\tau^{a}=U\tau^{b}U^{-1}E^{b}$ are the so called
 right electric fields and the function,
\BE
f\left(v\right) = \half \frac{\partial\ln N(v)}{\partial v}=
	\frac{I_{2}(2v)}{I_{1}(2v)}
	 \rightarrow  \left\{ \begin{array}{ll}
		 (v/2)(1-v^{2}/6)
		& v \ll 1   \\
	  1-3/(4v) & v \gg 1
		\end{array} \right.
\EE
is plotted in Fig.1 (solid line).
Due to the combined effect of the curvature of the SU(2) group manifold
and the finite width of the wave packet, the average of $E$ and
$U$ are not just their corresponding classical parameters as we would
expect for a Euclidean parameter  space. In the small width limit,
i.e. $v\rightarrow \infty$, $f(v)$ approaches unity,
a manifestation that locally
the group manifold has Euclidean metric. From this we already see that
the non-trivial geometry plays an important role here.
We may also study the quadratic fluctuations of $U$ and $E$.
We calculate
\BE
\langle E^{2} \rangle  =  \sum_{a}\langle E^{a}E^{a}\rangle=
	 \hbar^{2}\frac{3f}{8v}(v^{2}+w^{2})+
         \left(1-\frac{f}{2v}\right)E_{0}^{2}.
\EE
We shall later see
that the ensemble average of  $v$ is propotional to $ \hbar^{-1}$
and ensemble average of $ w$ is zero for $v \rightarrow \infty$.
So in the classical limit, i.e. $\hbar \rightarrow 0$, $\langle E^{2} \rangle$
goes to the classical result $E_{0}^{2}$.
The variance of electric fields is
\BE
(\Delta E)^{2} \equiv \sum_{a}\left(\langle E^{a}E^{a}\rangle - \langle E^{a}
	\rangle^{2}\right)=
	 \hbar^{2}\frac{3f}{8v}(v^{2}+w^{2})+
         (1-\frac{f}{2v}-f^{2})E_{0}^{2},
\EE
which goes to zero like $\frac{3}{8}\hbar^{2}v+\frac{E_{0}^{2}}{v}$
 in the classical limit.
The fluctuation of the vector potential $A$
is
\BE
(\Delta A)^{2} \equiv 8\langle 1-\half\tr (UU_{0}^{-1}) \rangle = 8(1-f).
\EE
It vanishes in the classical limit like $\hbar$.
We can check the uncertainty relation
$\Delta E \Delta A \ge 3(\hbar/2)$, which becomes an equality in
the vacuum. i.e. $T=0$ or $E_{0}=0$.
As we already said,
Gauss's law is satisfied on average, but it
is not conserved at the operator level. The magnitude of fluctuating
color charge at each lattice site is measured by
\BE
(\Delta G)^{2} \equiv \sum_{a}\left(\langle G^{a}G^{a}\rangle -\langle G^{a}
	\rangle^{2}\right),
\EE
where the second term vanishes. Inserting the expression of
the generator of time independent gauge transformation,
\BE
G^{a}=\sum_{i=1}^{3}\left(E^{a}({\bf n})-E^{a}_{R}({\bf n}-{\bf e}_{i})
	\right),
\EE
where the summation runs over the three positive
directions at one site, and ${\bf n}$
denotes the position of the site and ${\bf e}_{i}$ is the
unit vector in $i$th direction, we obtain
\BE
(\Delta G)^{2} = \sum_{i=1}^{6}(\Delta E)_{i}^{2},
\EE
where the summation now runs over all the links joining at one site.
Each link contributes its electric field fluctuation to
the local charge fluctuation.
The correlations between different links do not appear because
we have chosen a link separable wavefunction.

\subsection { Effective Hamiltonian }

The complete expression for the average total energy is
\BEA
      H_{sc} &=& H_{e} +H_{0} + H_{m}
	\equiv \langle H \rangle \nonumber \\
	  &=&
	 \sum_{l}\left[\frac{1}{2}(1-\frac{f_{l}}{2v_{l}})E_{l0}^{2}\right]+
\sum_{l}\left[\hbar^{2}\frac{3f_{l}}{16v_{l}}(v_{l}^{2}+w_{l}^{2})\right]+
	\nonumber \\
 	& & 4\sum_{p}\left(1-\half f_{p}\tr U_{p0}\right),
\EEA
where $f_{p}$ is the product of $f_{l}=f(v_{l})$ of the
four links that circumvent the plaquette $p$.
This semi-classical Hamiltonian consists of three parts.
First we have the modified electric energy ($H_{e}$) and
magnetic energy ($H_{m})$,
which result in a coupling between the $U_{l}$, $E_{l}$ and
the width parameters $v_{l}$ and $w_{l}$, and
thus modify the classical evolution of $U_{l}$ and $E_{l}$.
These modifications contain both
genuine quantum effects and lattice artifacts,
which are difficult to separate.
Besides the modified classical electric and magnetic energy
there is a zero point energy ($H_{0}$)
which is a consequence of the uncertainty relation.
It increases linearly with $v$
in the limit of large  $v$.

We can study thermodynamical properties of the system
described by $H_{sc}$.
Consider the system at equilibrium with temperature $T$.
The integration measures of $E$ and $U$ are the same
as in the classical case. The integration over $E$ is plagued by
Gauss's law which eliminates one third of the degrees of freedom.
The average magnetic energy is complicated by the interactions between
different links. They make it very difficult to obtain the exact solution
analytically. In the rest of this subsection,
let us simplify ourselves to consider
the Hamiltonian of
a single link and a single plaquette. In this free link and plaquette limit the
partition function can be factorized into three parts,
\BE
Z_{sc}=Z_{e}Z_{m}Z_{0},
\EE
with
\BEA
Z_{e} &=& \left( \frac{T}{2-f/v} \right)^{3/2}, \nonumber \\
Z_{m} &=& N(2f_{p}/T)\exp(-4/T), \nonumber \\
Z_{0} &=& \exp(-H_{0}/T), \nonumber
\EEA
where the subscripts $e,m,0$ denote contributions from
electric, magnetic and zero-point energy, respectively.
$N(x)$ is the normalization function defined in (28).
{}From the partition function we can
easily obtain the ensemble averaged energy
\BEA
\langle\langle H_{e}\rangle\rangle  &=& 3T/2, \nonumber \\
\langle\langle H_{m}\rangle\rangle &=& 4(1-f_{p}f(2f_{p}/T) ), \nonumber \\
\langle\langle H_{0}\rangle\rangle  &=& H_{0},
\EEA
where $\langle\langle \cdots \rangle\rangle$ denotes ensemble average.
We see that the electric energy is the same as that in the classical limit.
In the classical limit $f_{p}=1$ so $\langle\langle H_{m}
\rangle\rangle =4(1-f(2/T))$.
When $T \ll 2$, $\langle\langle H_{m}
\rangle\rangle$ takes the correct classical
limit $3T/2$ as demanded by the equipartition
theorem. At larger $T$ the energy temperature relation is distorted
by the compactness of the gauge group.
The free energy $F=-T\ln Z$ is obtained as
\BE
F(v)=	\frac{3T}{2}\ln \left((2-\frac{f}{v})\frac{1}{T}\right)+
	\frac{3\hbar^{2}}{16}fv+
	4-T\ln N\left(\frac{2f_{p}}{T}\right).
\EE
To see how the free energy looks like,
we  plot this function at $T=0.5$ and at $T=0$ in Fig2.
The two plots are very similar in shape, and
both have a first order phase transition.
At small $\hbar$ the free energy has two minima, one of which is at $v=0$.
The second minimum can be obtained in the limit of large $v$ by
minimizing the free energy, and the result is
\BE
v=2(16-7T)^{1/2}/\hbar.
\EE
When  $\hbar$ increases, this second minimum
rises until it fails to be a minimum at a critical
value of $\hbar$, where a first order
transition occurs.
The critical value
increases with temperature. From the plots
we see for $T=0$ it occurs at $\hbar\approx 1.73$ (the short dashed
line in Fig.2b) and $\hbar\approx 3$ for $T=0.5$
(the long dashed line in Fig.2a).
This indicates the underlying potential has a non-trival structure.
Vacuum energy is obtained from the free energy by setting $T=0$,
\BE
F_{0}=\frac{3\hbar^{2}}{16}fv+4(1-f^{4}).
\EE
At large $\hbar$, we only have one minimum $v=0$.
The wavefunction is so broad that it covers the whole
group manifold and we are in quantum physics region.
The corresponding energy is
$4$ (or $4/g^{2}a$ in original variables). This is coincident with the
leading order result of a perturbative calculation in the
strong coupling limit \cite{chin}.
In the other limit, $\hbar\rightarrow 0$, the wave packet can be
arbitrarily narrow, i.e. $v\rightarrow \infty$,
so it can probe the lowest point of
the potential well, and we are in the classical regime.
The energy density in this limit vanishes.

\subsection { Equations of Motion }

To obtain the semi-classical equations of motion we
calculate the expectation value of the time derivative:
\BE
\langle i \hbar\partial_{t}\rangle
		=\sum_{l} (- \hbar f{\dot w}_{l}+
                if \tr (E_{l0}{\dot U_{l0}}U_{l0}^{-1}) ).
\EE
Inserting this expression as well as
the one for $\langle H \rangle $  in (2) and performing the variation,
we obtain the equations of motion for each link (the subscript 0 is
suppressed and $f_{l}$ stands for $f(v_{l})$):
\BEA
\dot{E^{a}_{l}} & = &
\frac{i}{f_{l}} \sum_{p(l)} \tr\left( f_{p}\tau^{a}U_{p}\right)
		-\frac{3\hbar}{8}\frac{w_{l}}{v_{l}}E_{l}^{a}, \nonumber \\
\dot{U_{l}} & =
& \frac{i}{2f_{l}}\left(1-\frac{f_{l}}{2v_{l}}\right)E_{l}U_{l}, \nonumber \\
\dot{v_{l}} & = &
 \frac{3\hbar}{8} \frac{f_{l}}{f_{l}^{'}}\frac{w_{l}}{v_{l}},
			\nonumber \\
\dot{w_{l}} & = & \left(\frac{1}{f_{l}} - \frac{1}{2v_{l}}\right)E_{l}^{2}+
  \left(\frac{E_{l}^{2}}{4v_{l}}+\frac{2}{f_{l}}\sum_{p(l)}f_{p}\tr U_{p}-
		\frac{3}{16}\hbar^{2}(v_{l}+\frac{w_{l}^{2}}{v_{l}}) \right),
		\nonumber \\
	    &    & -\frac{f_{l}}{f_{l}^{'}}\left(\frac{E_{l}^{2}}{4v_{l}^{2}}+
		\frac{3}{16}\hbar^{2}(1-\frac{w_{l}^{2}}{v_{l}^{2}}) \right),
\EEA
where $f_{l}^{'}$ is the derivative of $f_{l}$ with respect to $v_{l}$,
the sum over $p(l)$ runs over all four plaquettes sharing the link $l$
and $U_{p}$ here is the ordered plaquette variable starting with link $U_{l}$.
We note that in the limit of small width or large $v$ the equations assume
the correct classical form,
\BEA
\dot{E_{l}^{a}} & = &
i \sum_{p(l)} \tr\left( \tau^{a}U_{p}\right), \nonumber \\
\dot{U_{l}} & = & iE_{l}U_{l}.
\EEA
Hence the equations (47) can be used to studied the quantum corrections to the
classical evolution (48).
These equations can also be studied in the limit where
$v$ is fixed and $w=0$, where the equations take the form
\BEA
\dot{E_{l}^{a}} & = &
 \frac{i}{f_{l}} \sum_{p(l)} \tr \left(f_{p}\tau^{a}U_{p}\right), \nonumber \\
\dot{U_{l}} & = & \frac{i}{2f_{l}}(1-\frac{f_{l}}{2v_{l}})E_{l}U_{l}.
\EEA
The fixed width parameter $v$ is to be treated as a variational parameter.

\subsection{ Classical Limit}

In order to be able to define a trajectory in phase space the wave packet
must be narrow compared to the total available phase space volume.
The restriction comes both from the link variables $U$ (coordinates)
and electric field variables $E$ (momenta).
To have a quantitative measure of how localized the wave packet is,
we define a quantity $\alpha(v)$ which is  the ratio of the volume of
the wave packet and the total group
volume, $2\pi^{2}$ for SU(2), i.e.:
\begin{eqnarray}
\alpha(v) = (1-f)^{3/2}  \rightarrow \left\{
		\begin{array}{ll}
		  1 & v
                          \rightarrow 0  \\
      		\left(\frac{3}{4v}\right)^{3/2} & v \rightarrow \infty.
		\end{array} \right.
\end{eqnarray}
The function $\alpha(v)$
is shown in Fig.1 (dotted line). If we request $\alpha(v) < 0.1 $, we
need $v >5$ and $\hbar < 1$ for $T \approx 1$.
The largest distance in momentum space is controlled by excitation
energy as $d_{E} \sim \sqrt{E}$ or
$d_{E} \sim T$. We require $\Delta E \ll d_{E}$. In the classical
limit ($v\rightarrow \infty$) this requires $ \hbar \ll T $,
which in the unscaled variables
is
\BE
\frac{\hbar}{Ta} =\frac{k_{max}}{\pi T} \ll 1.
\EE
where $k_{max}$ is the largest wave number that is possible to appear
on the lattice.
Physially this means that we shall only keep the long wavelength modes on
the lattice which is consistent to our intuition.

\section{Results and Discussions}

With the above semi-classical equations of motion we can follow
a trajectory in the space of gauge fields
for every given initial condition.
To make the results comparable to those of our calculations for
the classical gauge theory, we specify
the initial gauge configuration in the same way as in ref.[3].
We initially choose all electric field variables to be zero to
satisfy Gauss's law. The magnetic sector is initialized by chosing the
link variables $U_{l}=\cos(\rho/2)-i{\bf n}\cdot{\bf \tau} \sin(\rho/2)$
randomly and
the total energy is varied  by selecting $\delta$ which
limits the range of the parameter $\rho$ to $(0,2\pi \delta)$.
In addtion here we specify $\hbar$ and the width parameters.
Initially we choose identically on each link $v=20$ and $w=0$.
The results are given in Table 1.

To present our result we need to specify our controllable parameters.
One of these is $\hbar=4\pi\alpha_{s}$, which is a manifestation
of quantum effects. Another should be chosen to give a measure of the
excitation energy of the system.
A direct comparison between classical
and quantal systems with the same energy
is unphysical because of the
appearance of a  zero-point energy in the latter.
A convenient comparable parameter is the physical
temperature of an excited system.
The problem now is how to measure the temperature on the lattice.
For a weakly interacting system, we can extract the temperature
as follows.
The energy distribution for a small subsystem is measured,
which factorizes
into a phase space prefactor and an exponential $\exp(-E/T)$.
Dividing out the phase space factor we can simply identify the temperature as
the rate of exponential slope of the energy distribution.
But here this method cannot be used directly because
the lattice gauge system is a constrained and
strongly coupled system.
This problem will be discussed in
detail elsewhere.
Here we make the observation that the relative difference between the
electric part of the semi-classical Hamiltonian and that of the classical
one is $1/(2v)$ which is very small in the limit of large $v$.
As a consequence, the value of $\langle\langle H_{e}\rangle\rangle$,
the average electric energy per link,
at a certain  temperature $T$ of
the classical system and that of
the semi-classical system in the limit of large $v$
will be similar.
Neglecting the small difference,
we can use it as a measure of the excitation
level of the system,
and define parameter $T_{e} \equiv \langle\langle H_{e} \rangle\rangle $
as the "electric temperature" of the gauge field.


We have measured width parameters on the lattice. The results are
shown in Fig.3 for $\hbar=0.3$ and $T_{e}=0.9$, and in Table.1.
In Fig.3a and 3b,
the time evolution of the average values of $v$ and $w$
are shown. The average is taken over the lattice,
which is expected to coincide with the ensemble average if the lattice is
sufficiently large.
We find that these values are quickly
damped into their equilibrium positions, which are independent
of their initial value.
The final distributions of $v$ and $w$ over the lattice
are shown in Fig.3c and 3d respectively. The imaginary part of the
width parameter
$w$ obviously has zero
mean value.
The nonzero expectation value of $v$ depends mainly
on $\hbar$ as we can see from column 4 of Table 1.
We have obtained the least error fit
\BE
 \langle\langle v \rangle\rangle \approx 7 / \hbar.
\EE
The $T_{e}$ dependence is difficult to see because $T_{e}$ does not change
much between runs.
The $\hbar$ dependence roughly agrees with (44), which is found
for the system without
constraints and couplings by the standard thermodynamical method.

We have also measured the electric energy distribution
which is shown in Fig.4 for the same initial condition as in Fig.3.
The histogram is from the simulation. The solid curve is a fit,
\BE
f(H_{e}) \sim \sqrt{H_{e}}\exp(-H_{e}/0.6),
\EE
apart from a normalization constant. It has correct mean value
$\langle\langle H_{e} \rangle \rangle =0.9$ and looks very much like
a Boltzmann distribution. We note that
for a constrainted system it is possible
for the exponential in the energy distribution to have
the form $\exp(-\beta E/T)$ where $\beta$ can be different
from 1. So here we do not conclude the system
temperature is 0.6.

Now we come to the question which originally motivated this study, namely,
whether the quantum effects suppress or enhance the
chaotic behavior of a pure gauge system.
For this purpose we shall measure the maximal Lyapunov exponents $\lambda$
in our semi-classical formalism and compare them with the
ones we found for the classical lattice gauge theory at the same $T_{e}$.
The logrithmic scale of the divegence of trajectories
is shown by the solid line in Fig.5, where $\hbar=0.3$ and $T_{e}=0.9$.
For comparison, we also show in the same plot
the divegence of two corresponding
classical trajectories, with same $T_{e}$, by the dotted line.
The distance measure used here is the same as that
in ref.[3], i.e., the integrated absolute value of the local
difference in magnetic energy.
At early times, while the classical evolution smoothly increases the
distance between the two trajectories, there is a sudden jump of the
distance to a value independent of $D_{0}$ in the
semi-classical evolution, which originates from the couplings of
$U_{l}$ and $E_{l}$ to
the width parameters $v_{l}$ and $w_{l}$.
The latter oscillate and are damped strongly in the beginning as we
see in Fig.3.
Also at early times the fluctuations are large on the
logarithmical scale compared to the purely classical evolution.
These fluctuations are quantum effects.
At larger times the exponential increase of the distance
emerges from the fluctuating background. We see the rate of exponential
growth is larger than that of the classical one. The distance finally
reaches a maximum value. The maximal Lyapunov exponent is extracted
from the linear region in the plot. Due to
the large fluctuations this method is less accurate here than it is in the
classical case. The numbers obtained for $\lambda$ have errors of the order of
$\pm 0.02$. We performed
three runs for the system with the same $\hbar$ and $T_{e}$
and find that the statistical error is dominated by our measuring error.
The results for different initial conditions are
shown in column 3 of Table.1. At $\hbar=0.1$ the result deviates
very little from the classical result $\lambda{c}=T_{e}/3$.
The correction to $\lambda$ is represented by $\delta_{\lambda}$
(column 4) which is defined
as $\lambda(T_{e},\hbar) =  \lambda{c}(T_{e})(1+\delta\lambda)$.
$\delta\lambda$ depends both on $\hbar$ and $T_{e}$.
The relation is not easily extracted from the data due to the
lack of accuracy, but we note that the sign of
$\delta$ is always positive.
This suggests that quantum corrections
enhance the chaoticity of the classical gauge  system. The physical implication
is that including the quantum corrections,
the initial state of the gauge field will thermalize more rapidly.

We have also measured the time dependence of the distance between
two trajectories originating from the same point but one evolved
with the classical equations and the other with the variational quantum
equations. This distance also diverges exponentially,
because as soon as the quantum corrections
generate deviation between the trajectories the difference will
diverge with time because of the chaoticity of
our system. This leads to the question whether the
trajectory of the centroid of the wave packet
has any physical meaning. If we want to describe the quantum
corrections to a particular
classical trajectory exactly, then the method of Gaussian variational
states obviously fails.
However, if we are only interested in the ensemble averaged properties of the
system, e.g. the rate of the increase of the
coarse grained phase space volume, the approach remains useful.
 This is not too
serious a drawback because
even in the classical limit we cannot numerically follow
the exact evolution of a particular trajectory for a long time,
and we have to limit ourselves
to the study of average properties of the system.

In conclusion,
starting from a Gaussian ansatz we have derived a set of semi-classical
equations of motion for the lattice
SU(2) gauge field via a variational method.
These equations were used to study the quantum corrections
to the classical evolution. We find the quantum
effects enhance the chaoticity of the gauge field.

$Acknowledgements$:
We thank S. G. Martinyan and J. Rau for fruitful discussions.
This work was supported in part
by the U.S. Department of Energy (Grant No. DE-FG05-90ER40592) and
by a computing grant from the North Carolina Supercomputing Center.

\bigskip
{\noindent \Large \bf Table Captions}
\bigskip
\begin{itemize}
\item[Table.1] Results of numerical simulations.
\end{itemize}

\bigskip
{\noindent \Large \bf Figure Captions}
\bigskip
\begin{itemize}
\item[Fig.1] The solid line shows function $f(v)$ and
	the dotted line shows
	wave packet volume ratio $C(v)$ as a
	function of $v$ .
\item[Fig.2] Dependence of the free energy on $v$
	 for different values of $\hbar$ at
	 $T=0.5$ (a) and  $T=0$ (b). The sloid line is for
	 $\hbar=0.1$, the dotted line is for 1, the short dashed line
	 is for 1.732 and the long dashed line is for $\hbar=3$.
	 $v \rightarrow \infty$ corresponds to the classical limit of an
	 infinitely narrow wavepacket.
\item[Fig.3] Width properties for $T_{e}=0.9$ and $\hbar=0.3$:
	The equilibrium distribution of $v$ (a) and $w$ (b)
	and time evolution of the mean values of $v$ (c) and $w$ (d).
\item[Fig.4] Final electric energy distribution for $T_{e}=0.9$ and
 	$\hbar=0.3$.
\item[Fig.5] The solid line shows the
	divergence of the centroids for the two neighbouring
	Gaussian field configurations for $T_{e}=0.9$ and $\hbar=0.3$.
	The corresponding divergence for the two classical field
	configurations with the same $T_{e}$ is shown by the
	dotted line.
\end{itemize}

\newpage
\BC
Table.1
\EC

\begin{tabular}{@{\extracolsep{16mm}}c c c c c c}  \hline \hline
$\hbar$ & $T$ & $\lambda$ & $\delta\lambda$ & $\langle\langle v \rangle \rangle
$ & $E$ \\ \hline \hline
0.1 &    0.76 &    0.255 &   0.01 &    70 &      2.13   \\ \hline
0.3 &    0.64 &   0.24  &  0.12  &  23   &   2.48    \\ \hline
0.3 &    0.73 &   0.3  & 0.22   & 22    &  2.92    \\ \hline
0.3 &    0.85 &   0.34  & 0.2   & 21    &  3.31    \\ \hline
0.3 &    0.85 &   0.36  & 0.27   & 21    &  3.25    \\ \hline
0.3 &    0.85 &   0.34  & 0.2   & 22    &  3.25    \\ \hline
0.3 &    0.9  &   0.41  & 0.37   & 20    &  3.74    \\ \hline
0.7 &    0.5  &   0.19  & 0.16   & 9.0   &  3.24    \\ \hline
0.7 &    0.8  &   0.41  & 0.54   & 7.4   &  4.40    \\ \hline
\hline

\end{tabular}

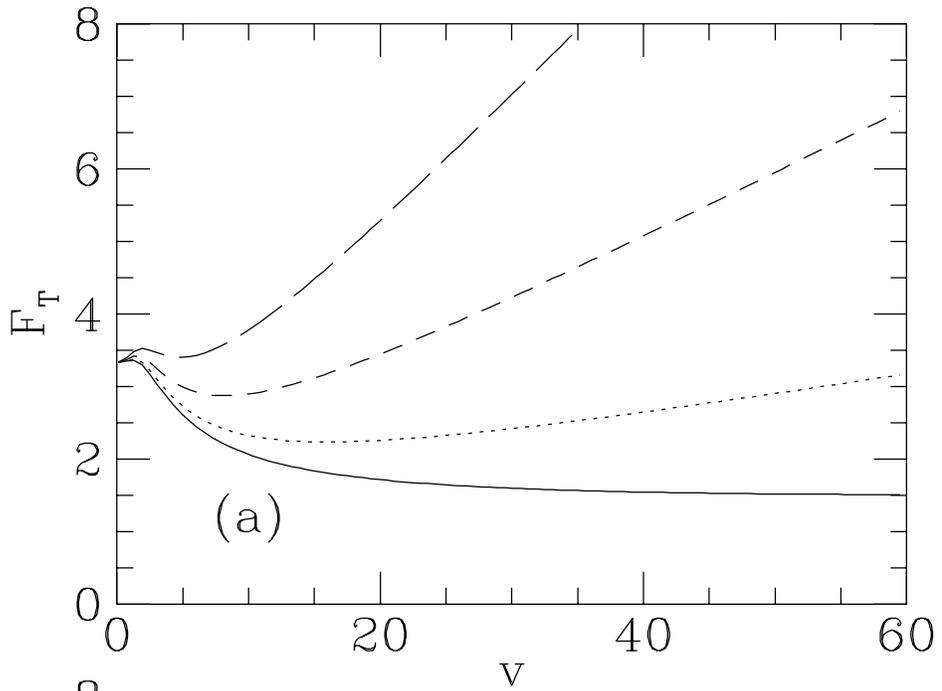
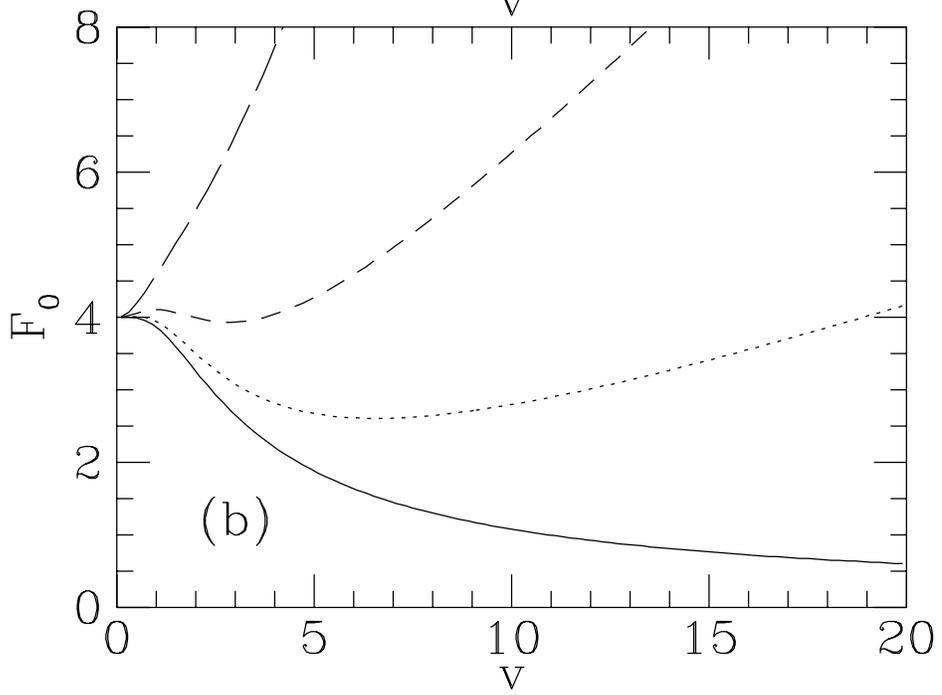
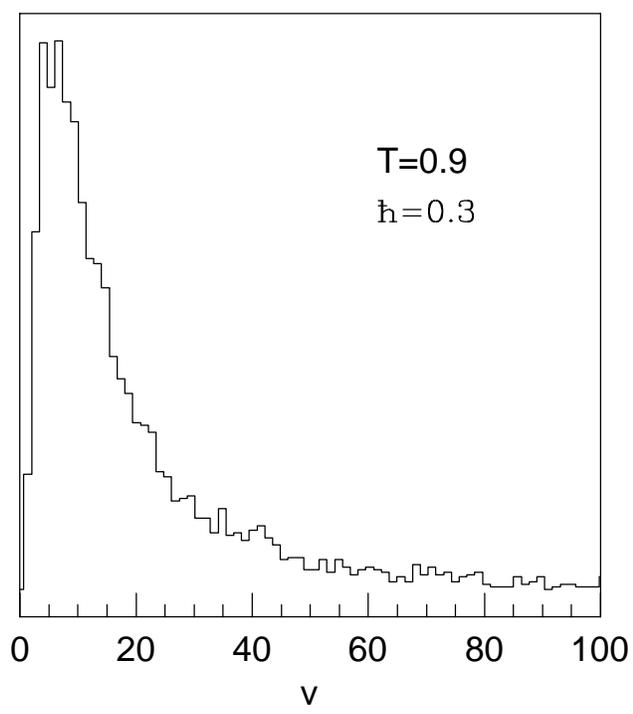
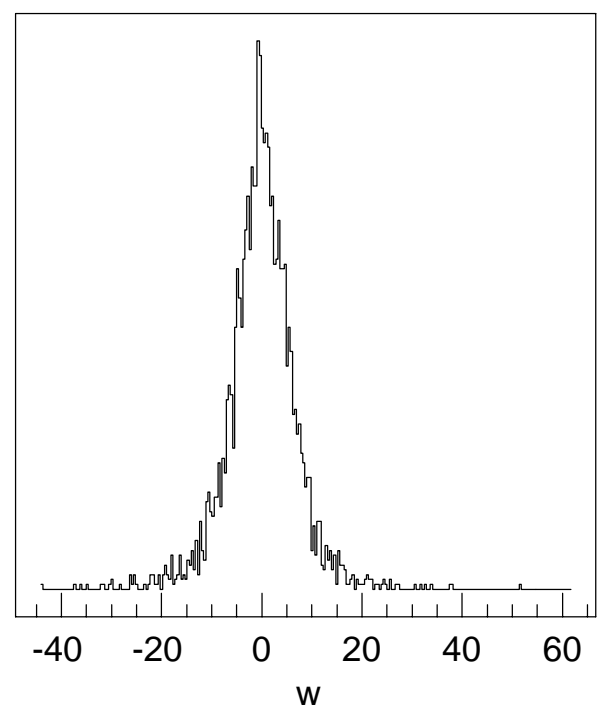
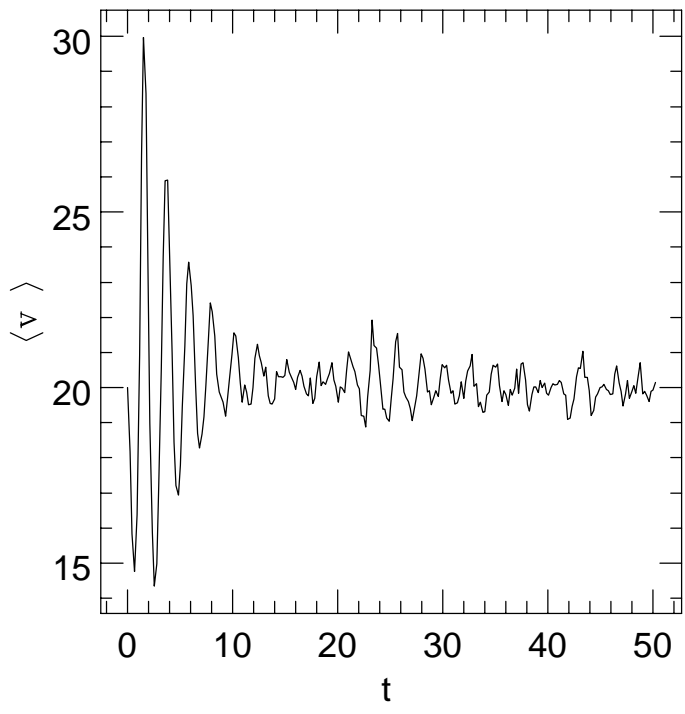
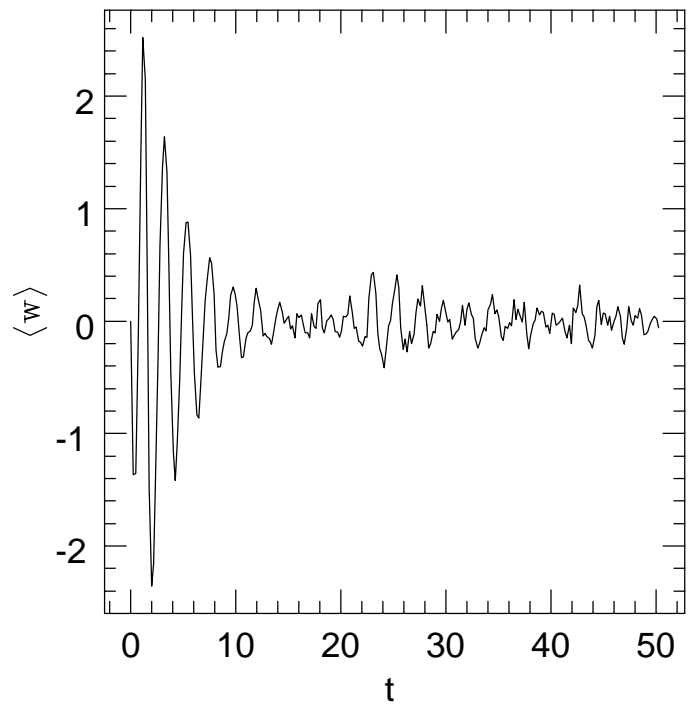
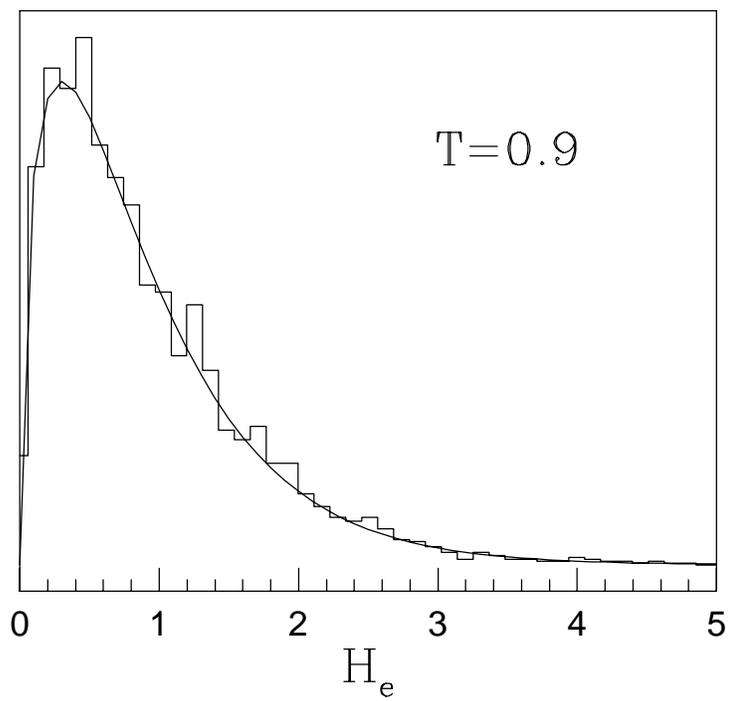
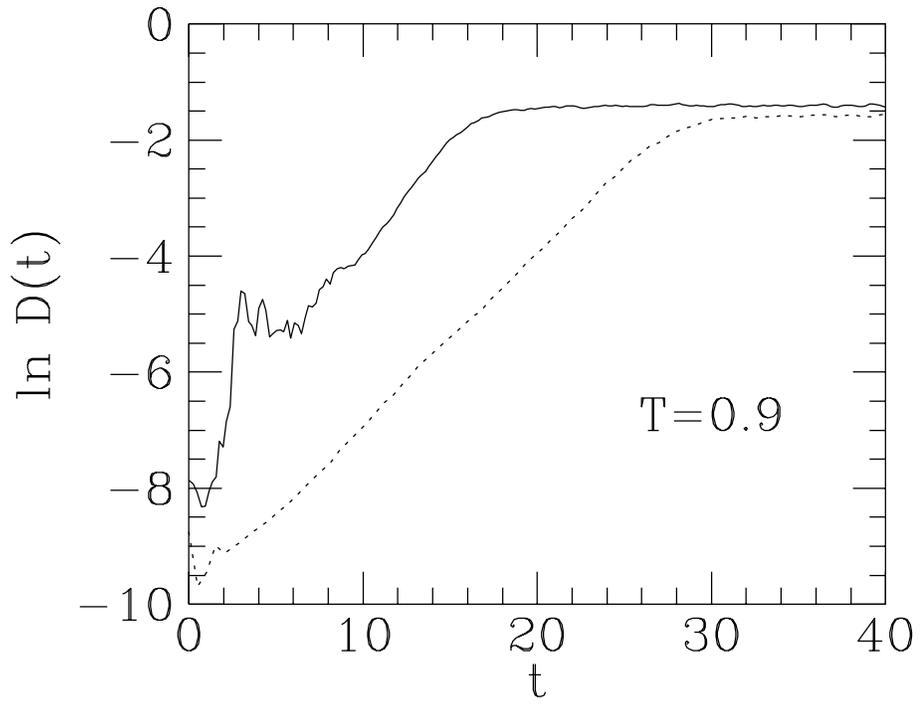
\end{document}